\input harvmac.tex


\def\unlockat{\catcode`\@=11}

\def\lockat{\catcode`\@=12}

\unlockat


\def\newsec#1{\global\advance\secno by1\message{(\the\secno. #1)}
\global\subsecno=0\global\subsubsecno=0
\global\deno=0\global\prono=0\global\teno=0\eqnres@t\noindent
{\bf\the\secno. #1}
\writetoca{{\secsym} {#1}}\par\nobreak\medskip\nobreak}
\global\newcount\subsecno \global\subsecno=0
\def\subsec#1{\global\advance\subsecno
by1\message{(\secsym\the\subsecno. #1)}
\ifnum\lastpenalty>9000\else\bigbreak\fi\global\subsubsecno=0
\global\deno=0\global\prono=0\global\teno=0
\noindent{\it\secsym\the\subsecno. #1}
\writetoca{\string\quad {\secsym\the\subsecno.} {#1}}
\par\nobreak\medskip\nobreak}
\global\newcount\subsubsecno \global\subsubsecno=0
\def\subsubsec#1{\global\advance\subsubsecno by1
\message{(\secsym\the\subsecno.\the\subsubsecno. #1)}
\ifnum\lastpenalty>9000\else\bigbreak\fi
\noindent\quad{\secsym\the\subsecno.\the\subsubsecno.}{#1}
\writetoca{\string\qquad{\secsym\the\subsecno.\the\subsubsecno.}{#1}}
\par\nobreak\medskip\nobreak}

\global\newcount\deno \global\deno=0
\def\de#1{\global\advance\deno by1
\message{(\bf Definition\quad\secsym\the\subsecno.\the\deno #1)}
\ifnum\lastpenalty>9000\else\bigbreak\fi
\noindent{\bf Definition\quad\secsym\the\subsecno.\the\deno}{#1}
\writetoca{\string\qquad{\secsym\the\subsecno.\the\deno}{#1}}}

\global\newcount\prono \global\prono=0
\def\pro#1{\global\advance\prono by1
\message{(\bf Proposition\quad\secsym\the\subsecno.\the\prono #1)}
\ifnum\lastpenalty>9000\else\bigbreak\fi
\noindent{\bf Proposition\quad\secsym\the\subsecno.\the\prono}{#1}
\writetoca{\string\qquad{\secsym\the\subsecno.\the\prono}{#1}}}

\global\newcount\teno \global\prono=0
\def\te#1{\global\advance\teno by1
\message{(\bf Theorem\quad\secsym\the\subsecno.\the\teno #1)}
\ifnum\lastpenalty>9000\else\bigbreak\fi
\noindent{\bf Theorem\quad\secsym\the\subsecno.\the\teno}{#1}
\writetoca{\string\qquad{\secsym\the\subsecno.\the\teno}{#1}}}
\def\subsubseclab#1{\DefWarn#1\xdef
#1{\noexpand\hyperref{}{subsubsection}%
{\secsym\the\subsecno.\the\subsubsecno}%
{\secsym\the\subsecno.\the\subsubsecno}}%
\writedef{#1\leftbracket#1}\wrlabeL{#1=#1}}

\lockat

\def\IB{\relax\hbox{$\inbar\kern-.3em{\rm B}$}}
\def\IC{\relax\hbox{$\inbar\kern-.3em{\rm C}$}}
\def\ID{\relax\hbox{$\inbar\kern-.3em{\rm D}$}}
\def\IE{\relax\hbox{$\inbar\kern-.3em{\rm E}$}}
\def\IF{\relax\hbox{$\inbar\kern-.3em{\rm F}$}}
\def\IG{\relax\hbox{$\inbar\kern-.3em{\rm G}$}}
\def\IGa{\relax\hbox{${\rm I}\kern-.18em\Gamma$}}
\def\IH{\relax{\rm I\kern-.18em H}}
\def\IK{\relax{\rm I\kern-.18em K}}
\def\IL{\relax{\rm I\kern-.18em L}}
\def\IP{\relax{\rm I\kern-.18em P}}
\def\IR{\relax{\rm I\kern-.18em R}}
\def\IZ{\relax\ifmmode\mathchoice
{\hbox{\cmss Z\kern-.4em Z}}{\hbox{\cmss Z\kern-.4em Z}}
{\lower.9pt\hbox{\cmsss Z\kern-.4em Z}}
{\lower1.2pt\hbox{\cmsss Z\kern-.4em Z}}\else{\cmss Z\kern-.4em
Z}\fi}

\def\II{\relax{\rm I\kern-.18em I}}


\def\CD {{\cal D}}
\def\CE {{\cal E}}
\def\CF {{\cal F}}

\def\CH {{\cal H}}
\def\CI {{\cal I}}

\def\CN {{\cal N}}

\def\CS {{\cal S}}

\def\CZ {{\cal Z}}



\def\Tr{{\rm Tr}}
\def\Id{{\rm Id}}

\def\Lie{{\rm Lie}}

\def\liet{{\underline{\bf t}}}

\def\inbar{\,\vrule height1.5ex width.4pt depth0pt}
\font\cmss=cmss10 \font\cmsss=cmss10 at 7pt


\font\manual=manfnt \def\dbend{\lower3.5pt\hbox{\manual\char127}}


\def\boxit#1{\vbox{\hrule\hbox{\vrule\kern8pt
\vbox{\hbox{\kern8pt}\hbox{\vbox{#1}}\hbox{\kern8pt}}
\kern8pt\vrule}\hrule}}
\def\mathboxit#1{\vbox{\hrule\hbox{\vrule\kern8pt\vbox{\kern8pt
\hbox{$\displaystyle #1$}\kern8pt}\kern8pt\vrule}\hrule}}

%
\lref\CMR{ For a review, see
S. Cordes, G. Moore, and S. Ramgoolam,
`` Lectures on 2D Yang Mills theory, Equivariant
Cohomology, and Topological String Theory,''
Lectures presented at the 1994 Les Houches Summer School
 ``Fluctuating Geometries in Statistical Mechanics and Field
Theory.''
and at the Trieste 1994 Spring school on superstrings.
hep-th/9411210, or see http://xxx.lanl.gov/lh94}
\lref\dnld{S. Donaldson, ``Anti self-dual Yang-Mills
connections over complex  algebraic surfaces and stable
vector bundles,'' Proc. Lond. Math. Soc,
{\bf 50} (1985)1}

\lref\DoKro{S.K.~ Donaldson and P.B.~ Kronheimer,
{\it The Geometry of Four-Manifolds},
Clarendon Press, Oxford, 1990.}
\lref\donii{
S. Donaldson, Duke Math. J. , {\bf 54} (1987) 231. }

\lref\gerasimov{A. Gerasimov, ``Localization in
GWZW and Verlinde formula,'' hepth/9305090}
\lref\kirwan{F.~Kirwan, ``Cohomology of quotients in symplectic
and algebraic geometry'', Math. Notes, Princeton University Press,
1985}
\lref\krwjffr{L.C.~Jeffrey, F.C.~Kirwan 
``Localization for nonabelian group actions'', alg-geom/9307001}
\lref\givental{A.B.~Givental,
``Equivariant Gromov - Witten Invariants'',
alg-geom/9603021}
\lref\prseg{Pressley and Segal, Loop Groups}
\lref\rade{J. Rade, ``Singular Yang-Mills fields. Local
theory I. '' J. reine ang. Math. , {\bf 452}(1994)111; {\it ibid}
{\bf 456}(1994)197; ``Singular Yang-Mills
fields-global theory,'' Intl. J. of Math. {\bf 5}(1994)491.}
\lref\segal{G. Segal, The definition of CFT}
\lref\sen{A. Sen,
hep-th/9402032, Dyon-Monopole bound states, selfdual harmonic
forms on the multimonopole moduli space and $SL(2,Z)$
invariance,'' }
\lref\shatashi{S. Shatashvili,
Theor. and Math. Physics, 71, 1987, p. 366}
\lref\thooft{G. 't Hooft , ``A property of electric and
magnetic flux in nonabelian gauge theories,''
Nucl.Phys.B153:141,1979}
\lref\vafa{C. Vafa, ``Conformal theories and punctured
surfaces,'' Phys.Lett.199B:195,1987 }
\lref\vrlsq{E. Verlinde and H. Verlinde,
``Conformal Field Theory and Geometric Quantization,''
in {\it Strings'89},Proceedings
of the Trieste Spring School on Superstrings,
3-14 April 1989, M. Green, et. al. Eds. World
Scientific, 1990}

\lref\mwxllvrld{E. Verlinde, ``Global Aspects of
Electric-Magnetic Duality,'' hep-th/9506011}

\lref\wrdhd{R. Ward, Nucl. Phys. {\bf B236}(1984)381}
\lref\ward{Ward and Wells, {\it Twistor Geometry and
Field Theory}, CUP }

\lref\witdyn{E. Witten, ``String theory dynamics
in various dimensions,''
hep-th/9503124, Nucl. Phys. {\bf B} 443 (1995) 85-126}
\lref\WitDonagi{R.~ Donagi, E.~ Witten,
``Supersymmetric Yang-Mills Theory and
Integrable Systems'', hep-th/9510101, Nucl. Phys.{\bf B}460 (1996)
299-334}
\lref\Witfeb{E.~ Witten, ``Supersymmetric Yang-Mills Theory On A
Four-Manifold,''  hep-th/9403195; J. Math. Phys. {\bf 35} (1994) 5101.}
\lref\Witr{E.~ Witten, ``Introduction to Cohomological Field
Theories",
Lectures at Workshop on Topological Methods in Physics, Trieste,
Italy,
Jun 11-25, 1990, Int. J. Mod. Phys. {\bf A6} (1991) 2775.}
\lref\Witgrav{E.~ Witten, ``Topological Gravity'',
Phys.Lett.206B:601, 1988}
\lref\witaffl{I. ~ Affleck, J.A.~ Harvey and E.~ Witten,
	``Instantons and (Super)Symmetry Breaking
	in $2+1$ Dimensions'', Nucl. Phys. {\bf B}206 (1982) 413}
\lref\wittabl{E.~ Witten,  ``On $S$-Duality in Abelian Gauge
Theory,''
hep-th/9505186; Selecta Mathematica {\bf 1} (1995) 383}
\lref\wittgr{E.~ Witten, ``The Verlinde Algebra And The Cohomology Of
The Grassmannian'',  hep-th/9312104}
\lref\wittenwzw{E. Witten, ``Nonabelian bosonization in
two dimensions,'' Commun. Math. Phys. {\bf 92} (1984)455 }
\lref\witgrsm{E. Witten, ``Quantum field theory,
grassmannians and algebraic curves,'' Commun.Math.Phys.113:529,1988}
\lref\wittjones{E. Witten, ``Quantum field theory and the Jones
polynomial,'' Commun.  Math. Phys., 121 (1989) 351. }
\lref\witttft{E.~ Witten, ``Topological Quantum Field Theory",
Commun. Math. Phys. {\bf 117} (1988) 353.}
\lref\wittmon{E.~ Witten, ``Monopoles and Four-Manifolds'',
hep-th/9411102}
\lref\Witdgt{ E.~ Witten, ``On Quantum gauge theories in two
dimensions,''
Commun. Math. Phys. {\bf  141}  (1991) 153}
\lref\witrevis{E.~ Witten,
 ``Two dimensional gauge
theories revisited'', hep-th/9204083;
J. Geom. Phys. 9 (1992) 303-368}
\lref\Witgenus{E.~ Witten, ``Elliptic Genera and Quantum Field
Theory'',
Comm. Math. Phys. 109(1987) 525. }
\lref\OldZT{E. Witten, ``New Issues in Manifolds of SU(3) Holonomy,''
{\it Nucl. Phys.} {\bf B268} (1986) 79 \semi
J. Distler and B. Greene, ``Aspects of (2,0) String
Compactifications,''
{\it Nucl. Phys.} {\bf B304} (1988) 1 \semi
B. Greene, ``Superconformal Compactifications in Weighted Projective
Space,'' {\it Comm. Math. Phys.} {\bf 130} (1990) 335.}

\lref\bagger{E.~ Witten and J. Bagger, Phys. Lett.
{\bf 115B}(1982) 202}

\lref\witcurrent{E.~ Witten,``Global Aspects of Current Algebra'',
Nucl.Phys.B223 (1983) 422\semi
``Current Algebra, Baryons and Quark Confinement'', Nucl.Phys. B223
(1993)
433}
\lref\Wittreiman{S.B. Treiman,
E. Witten, R. Jackiw, B. Zumino, ``Current Algebra and
Anomalies'', Singapore, Singapore: World Scientific ( 1985) }
\lref\Witgravanom{L. Alvarez-Gaume, E.~ Witten, ``Gravitational
Anomalies'',
Nucl.Phys.B234:269,1984. }

\lref\CHSW{P.~Candelas, G.~Horowitz, A.~Strominger and E.~Witten,
``Vacuum Configurations for Superstrings,'' {\it Nucl. Phys.} {\bf
B258} (1985) 46.}

\lref\AandB{E.~Witten, in ``Proceedings of the Conference on Mirror
Symmetry",
MSRI (1991).}

\lref\phases{E.~Witten, ``Phases of N=2 Theories in Two Dimensions",
Nucl. Phys. {\bf B403} (1993) 159, hep-th/9301042}
\lref\WitKachru{S.~Kachru and E.~Witten, ``Computing The Complete
Massless
Spectrum Of A Landau-Ginzburg Orbifold,''
Nucl. Phys. {\bf B407} (1993) 637, hep-th/9307038}

\lref\WitMin{E.~Witten,
``On the Landau-Ginzburg Description of N=2 Minimal Models,''
IASSNS-HEP-93/10, hep-th/9304026.}

\lref\wittenwzw{E. Witten, ``Nonabelian bosonization in
two dimensions,'' Commun. Math. Phys. {\bf 92} (1984)455 }
\lref\grssmm{E. Witten, ``Quantum field theory,
grassmannians and algebraic curves,'' Commun.Math.Phys.113:529,1988}
\lref\wittjones{E. Witten, ``Quantum field theory and the Jones
polynomial,'' Commun.  Math. Phys., 121 (1989) 351. }
\lref\wittentft{E.~ Witten, ``Topological Quantum Field Theory",
Commun. Math. Phys. {\bf 117} (1988) 353.}
\lref\Witr{E.~ Witten, ``Introduction to Cohomological Field
Theories",
Lectures at Workshop on Topological Methods in Physics, Trieste,
Italy,
Jun 11-25, 1990, Int. J. Mod. Phys. {\bf A6} (1991) 2775.}
\lref\wittabl{E. Witten,  ``On S-Duality in Abelian Gauge Theory,''
hep-th/9505186}
\lref\witbound{E.~Witten, ``Bound States Of Strings And $p$-Branes'',
hep-th/9510135
Nucl. Phys. {\bf B}460 (1996) 335-350}
\lref\witconst{E.~Witten, ``Constraints on supersymmetry breaking'',
Nucl. Phys. {\bf B}202 (1982) 253}


\lref\seiken{K. Intriligator, N. Seiberg,
``Mirror Symmetry in Three Dimensional Gauge Theories'',
hep-th/9607207, Phys.Lett. B387 (1996) 513}
\lref\douglas{M.R. Douglas, ``Enhanced Gauge
Symmetry in M(atrix) Theory,'' hep-th/9612126}
\lref\hs{J.A. Harvey and A. Strominger,
``The heterotic string is a soliton,''
hep-th/9504047}
\lref\hm{ J.A.~Harvey, G.~Moore,
``On the algebras of BPS states'', hep-th/9609017}
\lref\zt{O.~Aharony, M.~Berkooz, N.~Seiberg, ``Light-cone description
of $(2,0)$ superconformal theories in six dimensions'',
hep-th/9712117}
\lref\sen{A. Sen, `` String- String Duality Conjecture In Six
Dimensions And
Charged Solitonic Strings'',  hep-th/9504027}
\lref\KN{P.~Kronheimer and H.~Nakajima,  ``Yang-Mills instantons
on ALE gravitational instantons,''  Math. Ann.
{\bf 288}(1990)263}
\lref\nakajima{H.~Nakajima, ``Homology of moduli
spaces of instantons on ALE Spaces. I'' J. Diff. Geom.
{\bf 40}(1990) 105; ``Instantons on ALE spaces,
quiver varieties, and Kac-Moody algebras,'' Duke. Math. J. {\bf 76}
(1994)
365-416\semi
``Gauge theory on resolutions of simple singularities
and affine Lie algebras,'' Inter. Math. Res. Notices (1994), 61-74}
\lref\nakheis{H.~Nakajima,
``Lectures on Hilbert schemes of points on surfaces'', H.~Nakajima's
homepage}
\lref\vw{C.~Vafa, E.~Witten, ``A strong coupling test of
$S$-duality'', hep-th/9408074;
Nucl. Phys. {\bf B} 431 (1994) 3-77}
\lref\grojn{I. Grojnowski, ``Instantons and
affine algebras I: the Hilbert scheme and
vertex operators,'' alg-geom/9506020}
\lref\gr{G.~Gibbons, P.~Rychenkova ``hyperkahler quotient
construction
of BPS Monopole moduli space'', hep-th/9608085}
\lref\dvafa{C.~Vafa, ``Instantons on D-branes'', hep-th/9512078,
Nucl. Phys. B463 (1996) 435-442}
\lref\atbott{M.~Atiyah, R.~Bott, ``The Moment Map And
Equivariant Cohomology'', Topology {\bf 23} (1984) 1-28}
\lref\atbotti{M.~Atiyah, R.~Bott, ``The Yang-Mills Equations Over
Riemann Surfaces'', Phil. Trans. R.Soc. London A {\bf 308}, 523-615
(1982)}
\lref\cs{L.~Baulieu, A.~Losev, N.~Nekrasov,
``Chern-Simons and Twisted Supersymmetry in Higher Dimensions'',
hep-th/9707174, to appear in Nucl. Phys. B}
\lref\nikitafive{N.~Nekrasov, ``Five Dimensional Gauge Theories
and Relativistic Integrable Systems'', hep-th/9609219}
\lref\marty{M.~Claudson, M.B.~Halpern, `Supersymmetric ground state wave
functions'', 
Nucl. Phys. {\bf B} 250(1985)  689}
\lref\polchin{S. Chaudhuri, C. Johnson, and J. Polchinski,
``Notes on D-branes,'' hep-th/9602052; J. Polchinski,
``TASI Lectures on D-branes,'' hep-th/9611050}
\lref\ginz{V.~Ginzburg, R.~Besrukavnikov, to appear}
\lref\grn{M.~Green, M.~Gutperle, ``$D$-particle bound states and the
$D$-instanton measure'',
hep-th/9711107}
\lref\sav{S.~Sethi, M.~Stern, ``$D$-brane bound states redux,''
hep-th/9705046}
\lref\pyi{P.~Yi,
``Witten Index and Threshold Bound States of D-Branes''
hep-th/9704098,  Nucl. Phys. {\bf B} 505 (1997) 307-318}
\lref\higgs{G.~Moore, N.~Nekrasov, S.~Shatashvili, ``Integrating over
Higgs Branches'',
hep-th/9712241}
\lref\kato{S. Hirano, M. Kato, ``Topological Matrix Model',  
hep-th/9708039, Prog.Theor.Phys. 98 (1997) 1371.}

\lref\horostrom{G.T. Horowitz and A. Strominger,
``Black strings and $p$-branes,''
Nucl. Phys. {\bf B360}(1991) 197}
\lref\nicolai{W.~Krauth, H.~Nicolai, M.~Staudacher,
``Monte Carlo Approach to M-theory'', hep-th/9803117}
\lref\manin{Yu.~Manin, ``Generating functions in algebraic
geometry and sums over trees'', alg-geom/9407005}
\lref\estring{J.A.~Minahan, D.~Nemeschansky, C.~Vafa, N. P.~Warner,
``E-Strings and $\CN=4$ Topological Yang-Mills Theories'',
hep-th/9802168}
\lref\townsend{P. Townsend, ``The eleven dimensional supermembrane
revisited,'' hep-th/9501068}
\lref\porrati{M.~Porrati, A.~Rozenberg,
``Bound States at Threshold in Supersymmetric Quantum Mechanics'',
hep-th/9708119}
\lref\ikkt{N. Ishibashi, H. Kawai, Y. Kitazawa, and A. Tsuchiya,
``A large $N$ reduced model as superstring,'' hep-th/9612115;
Nucl. Phys. {\bf B498}(1997)467.}

\Title{ \vbox{\baselineskip12pt\hbox{hep-th/9803265}
\hbox{CERN-TH/98-83}
\hbox{HUTP- 98/A008}
\hbox{ITEP-TH.8/98}
\hbox{NSF-ITP-98-031}
\hbox{YCTP-P6/98}
}}
{\vbox{
 \centerline{$D$-particle bound states and}
 \centerline{generalized instantons}}}
\medskip
\centerline{Gregory Moore, $^1$
Nikita Nekrasov, $^{2,3}$ and Samson Shatashvili
$^{1,4}$\footnote{*}{On
leave of
absence from St. Petersburg Branch of Steklov Mathematical Institute,
Fontanka,
St.
Petersburg,
Russia.}}

\vskip 0.5cm
\centerline{$^{1}$ Department of Physics, Yale University,
New Haven, CT  06520, Box 208120, USA}
\centerline{$^{2}$ Institute of Theoretical and Experimental
Physics,
117259, Moscow, Russia}
\centerline{$^{3}$ Lyman Laboratory of Physics,
Harvard University, Cambridge, MA 02138, USA}
\centerline{$^{4}$ Theory Division, CERN, CH-1211, Geneve 23,
Switzerland}
\vskip 0.1cm
\centerline{moore@castalia.physics.yale.edu}
\centerline{nikita@string.harvard.edu }
\centerline{samson@euler.physics.yale.edu}

\medskip
\noindent
We compute the principal contribution to the index in the supersymmetric
quantum mechanical systems which are
obtained by reduction to $0+1$ dimensions
of $\CN=1$, $D=4,6,10$ super-Yang-Mills theories with gauge group $SU(N)$.
The results are:  ${1\over{N^{2}}}$ for  $D=4,6$,
$\sum_{d \vert N} {1\over{d^{2}}}$
for $D=10$. We also discuss the $D=3$ case.

\Date{March 31, 1998}

\newsec{Introduction }

The existence of $M$-theory
depends crucially on the existence within
type-$IIA$ string theory of a tower of massive
BPS particles electrically charged with
respect to the RR $1$-form.
These particles, originally described
as black holes in $IIA$ supergravity \horostrom,
can be interpreted as Kaluza-Klein particles of eleven-dimensional $M$-theory
compactified
on a circle  \townsend\witdyn.
Later, these particles were identified with
``$D0$-branes'' \polchin. In the $D$-brane
formulation it becomes clear that in certain
energy regimes the
dynamics of $N$ such particles
can be described by the supersymmetric quantum mechanics of $N \times N$
Hermitian matrices obtained from  dimensional reduction of $\CN=1$, $D=10$
super-Yang-Mills  theory \witbound (the quantum mechanical
model was originally
studied in \marty). The existence of the
$M$-theoretic
Kaluza-Klein tower of states is equivalent to the
statement that this quantum mechanics has exactly
one bound state for each $N$. Consequently,
proving the existence of these bound states has
been the focus of several recent papers of which
\sav\grn\porrati\ are the most relevant to the present
work. In particular, we note that the existence of
the bound state  in the case of $N=2$ was
proven in \sav, but the case $N>2$ remains open.
The results of the present paper will help complete
the proof for all $N$.

The existence of bound states in susy
quantum mechanics can be detected by
computing  the Witten index:
\eqn\wtin{{\rm lim}_{\beta \to \infty} {\Tr}_{\CH} (-1)^{F}
e^{-\beta  H} =
N_{B} - N_{F}}
where $N_{B,F}$ are the numbers of bosonic and fermionic
zero eigen-states of the Hamiltonian $H$ respectively.
The expression ${\Tr}_{\CH} (-1)^{F}  e^{-\beta  H}$ is
$\beta$-independent
in   theories with a discrete spectrum, but may be rather complicated
if the spectrum is continuous. In fact, the densities of fermionic
and bosonic
eigen-states may differ, leading to nontrivial $\beta$-dependence.
Nevertheless,
  supersymmetry allows us to relate the index of interest
to the easier-to-access quantity:
\eqn\es{{\rm lim}_{\beta \to 0} {\Tr}_{\CH} (-1)^{F}  e^{-\beta  H}
\quad . }
In the case of  the quantum mechanics of $N$ $D0$-branes
\es\ can be expressed very explicitly as a matrix integral
\eqn\mmint{{1\over{{\rm Vol}(G)}} \int d^{10} X d^{16} \Psi e^{-S}}
where $S$ is the reduction to zero dimensions of the action of the
$\CN=1$ $d=10$ super-Yang-Mills theory with the gauge group
$G = SU(N)/{\IZ}_{N}$.
More generally, we are aiming at computing the integral
\eqn\innt{
I_D(N)\equiv
\left({{\pi}\over{g}}\right)^{{(N^{2}-1)(D-3)}\over{2}} 
{1\over{{\rm Vol}(G)}} 
\int d^{D} X d^{2^{D/2-1}} \Psi
e^{-S}}
for $D= 3+1, 5+1, 9+1$ respectively, where
\eqn\actn{S = {1\over{g}} \left( {1\over{4}} \sum_{\mu, \nu = 1,
\ldots , D}
{\Tr} [ X_{\mu}, X_{\nu} ]^{2} + {i \over{2}}\sum_{\mu = 1}^{D} {\Tr}
(\bar\Psi \Gamma^{\mu}
[ X_{\mu}, \Psi] )\right),}
and
$\Gamma^{\mu}$ are the Clifford matrices for $Spin(D)$.
The integrals \mmint\innt\ are not the full contribution
to the Witten index (indeed, as we will see, they are
not integral).
The difference (also called the boundary term)
\eqn\dfnce{{\rm lim}_{\beta \to \infty} {\Tr}_{\CH} (-1)^{F}
e^{-\beta  H}-
{\rm lim}_{\beta \to 0} {\Tr}_{\CH} (-1)^{F}  e^{-\beta  H} =
\int_{0}^{\infty} d{\beta}
{d \over  d \beta} {\Tr}_{\CH} (-1)^{F}  e^{-\beta  H}}
may be analysed separately and is   beyond the scope of
this paper. See \sav\grn\porrati\ for further discussion.

The paper is organized as follows. In
section $2$ we reinterpret
the integrals \mmint\innt\ as those appearing
in the CohFT
approach to the studies of the
moduli space of susy gauge configurations,
reduced to $0$ dimensions.
\foot{``CohFT'' $=$ ``Cohomological field theory.''}
The susy gauge  configurations obey flatness, instanton
and complexified (or octonionic)
instanton equations in $3+1$, $5+1$ and $9+1$ cases respectively.
(Quantum mechanics on the moduli spaces of such susy
gauge configurations on compact manifolds was studied recently
in \cs.)

In  section $3$ we deform
the integral using the global
symmetries of the equations. The symmetry
groups   are
$Spin(2)$, $Spin(4)$ and $Spin(6)$
(or $Spin(7)$) in $D=4,6,10$, respectively. We simplify the deformed integrals
by the method of
``integrating out  BRST quartets''
and get contour integrals over the
eigenvalues of one of the matrices, denoted $\phi$
below. This brings the integrals to the form given
in equations $(3.6)$, $(3.7)$, and $(3.8)$ below for
the cases
$D=10,6,4$. The method used to arrive at these
expressions is a direct extension of  methods
we used to integrate over Higgs branches in \higgs.

The expressions $(3.6) ,  (3.7) , (3.8)$ are one of
the main results of this paper. Nevertheless, we
must note at the outset that the result is incomplete.
As Lebesgue integrals these expressions do not
make sense. Rather, they should be regarded as contour
integrals, which do make sense once a prescription is adopted for picking up
the poles of the integrand. We are confident that a more careful implementation
of the quartet mechanism will lead to a definite pole prescription. In this
paper we will take
the pragmatic route and simply find a pole prescription
which gives the desired answer.
In particular, in section $4$ we perform an  explicit evaluation
 of \innt\ for   $G=SU(2), SU(3)$.
In sections $5,6,7$ we evaluate
the integrals for the general case $G=SU(N)$. Each case,
$D=4,6,10$,  requires a different trick in order to carry
out the intricate sum over poles. In the $D=3+1$ case we
use an identity familiar from bosonization in two-dimensions.
In the $D=5+1$ case we use fixed-point techniques
for a certain torus action on the Hilbert
scheme of $N$ points on $\IC^{2}$. In section $7$ we
deform the octonionic instanton equations and
reduce the
$D=9+1$ case to a sum over answers for $D=3+1$ with
the sum running over all  possible
unbroken gauge groups of  $\CN=4$ super-Yang-Mills theory broken down to
$\CN=1$ by   mass terms.
This final reduction  leads to an answer for the index
 computation, predicted by
M.~Green and M.~Gutperle in  \grn,
building on the work of \sav.

Finally, in section $8$ we relate our
computations to the partition
functions of the SYM theories on $T^{4}, K3, T^{3}$ and discusses some
subtleties of the latter case.

As our paper was nearing completion  a
related paper appeared  \nicolai. This paper
describes a complementary (numerical) approach to the evaluation
of the integrals $I_{D}(N)$ and in particular evaluates the integral for
the
$SU(3)$ case.
Also, the paper \porrati\   studied the mass deformations
of the quantum mechanical problems we consider here, for
the case when
$N$ is prime.
It would be interesting to understand better the relation to these 
works. It was brought to our attention that the CohFT reformulation
of IKKT model has been also considered in \kato.

\newsec{CohFT reinterpretation}

To map to CohFT formalism we choose two matrices, say $X_{D}$ and
$X_{D-1}$,
and arrange them into a complex matrix $\phi$:
$$
\phi = X_{D-1} + i X_{D}
$$
The rest of the matrices can be written
 as $B_{j} = X_{2j-1} + i X_{2j}$
for $j=1, \ldots , D/2-1$.
Sometimes we simply denote them as ${\bf X} = \{ X_{a}, a=1, \ldots,
D-2 \}$.
We also rearrange the fermions:
$\Psi \to \Psi_{a} = (\psi_{j}, \psi_{j}^{\dagger}), \vec\chi, \eta$
and add   bosonic auxiliary fields $\vec H$.
Then we rewrite the bosonic part of the action as:

\eqn\rwrac{S = {1\over{16 g}} {\Tr} [\phi, \bar\phi]^{2} - i {\Tr}
\vec\CE ({\bf X}) \vec H + g {\Tr} \vec H^{2} -
{1\over{4g}} \sum_{a=1}^{D-2}  {\Tr} \vert [ X_{a}, \phi ]\vert^{2}}
where the ``equations'' $\vec\CE$ are:

\eqna\fsten
$$
\eqalignno{
D=4: \qquad &   \vec\CE = [B_{1}, B_{1}^{\dagger}] & \fsten a \cr
D=6: \qquad &   \vec\CE = \left( [B_{1}, B_{1}^{\dagger}] +
[B_{2}, B_{2}^{\dagger}],
[B_{1}, B_{2}], [B_{2}^{\dagger}, B_{1}^{\dagger}] \right)
\qquad\quad\qquad\quad\qquad\quad\qquad\quad  \fsten b \cr
D=10: \qquad &   \vec \CE = \left( [B_{i}, B_{j}] + \half \epsilon_{ijkl}
[B_{k}^{\dagger}, B_{l}^{\dagger}], i < j, \quad
\sum_{i} [B_{i}, B_{i}^{\dagger} ] \right)
\qquad\quad\qquad\quad\qquad\quad\fsten c \cr}$$

It is worth noting that one can also write the equations
\fsten{b}
as a three-vector $ \CE_{A} =
[X_{A}, X_{4} ] + \half \varepsilon_{ABC}
 [X_{B}, X_{C}]$, $A=1,2,3$.
Similarly, we can also write the equations \fsten{c}
as a
seven-vector:
$\CE_{\bf A} = [X_{\bf A}, X_{8} ] +
\half c_{\bf ABC} [X_{\bf B}, X_{\bf
C}]$ using octonionic
structure constants: $\bf A = 1, \ldots, 7$.

The integral \innt\ has the following important nilpotent symmetry:
\eqn\ssy{\eqalign{Q X_{a} = \Psi_{a} \quad & Q \Psi_{a} = [\phi ,
X_{a} ]\cr
Q \vec \chi = \vec H \quad & Q \vec H = [\phi, \vec \chi] \cr
Q \bar\phi = \eta \quad & Q \eta = [\phi, \bar\phi] \cr
 Q & \phi = 0\cr }}

In fact, the action \rwrac\ together with fermions can be represented
as:
\eqn\rwraci{S = Q\left( {\Tr} {1\over{16g}} \eta[\phi, \bar\phi] - i {\Tr} \vec
\chi \cdot \vec \CE + g {\Tr} \vec \chi \cdot \vec H +
{1\over{4g}} \sum_{a=1}^{D-2} {\Tr} \Psi_{a} [ X_{a}, \bar \phi] \right)}

As usual, there is a ghost charge. It is equal to $+2$ for $\phi$,
$+1$ for $\Psi_{a}$, $0$ for
$\vec H$, $X_{a}$, $-1$ for $\vec \chi, \eta$ and $-2$ for
$\bar\phi$.

All the bosonic fields except $\phi$ are paired with the fermions.
Therefore,  in order to fix the normalization of the integral one
need only fix the measure $D\phi$ on the Lie algebra of $G$. Since $\Lie G$ is
a simple Lie algebra, there is a unique Killing form up to a constant multiple.
This form determines
 the measure both on the Lie algebra and
on the group $G$. The measure
$$
{{D \phi}\over{{\rm Vol}(G)}}
$$
is thus independent of the choice of the Killing form.
However, the measure depends on whether the gauge group contains
the center or not. We ought to  use the measure normalized against
$G  = SU(N)/{\IZ}_{N}$, since it is $G$ which is the
actual gauge group of the problem. When we
reduce the computation
to an integral over the Lie algebra of the
maximal torus $T \subset
SU(N)$ the measure $\CD \phi$ will be
 normalized in such a way that the
measure on $T$ obtained by the exponential map integrates to one.
Therefore there is an extra factor
$\# Z$ in front of the integral since in passing to the measure on
$\liet$
we get as a factor a volume of the generic adjoint orbit:
\eqn\grpvol{
{{{\rm Vol} (G / (T/Z))}\over{{\rm Vol}(G)}} = {{\# Z}\over{{\rm
Vol}(T)}}.
}
Finally, upon eliminating the auxilliary fields $\vec H$ by taking the 
Gaussian  integral the extra factor 
$\left( {{\pi}\over{g}} \right)^{{{(D-3)(N^{2}-1)}\over{2}}}$ 
appears. It is related to the $\beta$-dependent factor appearing
in the index computation \sav.
 
\newsec{Global symmetries and deformation}

The global symmetries are

\item{4}: $K = Spin(2)$, $\vec\CE \in {\bf 1}$;

\item{6}: $K= Spin(4)$, $\vec \CE \in {\bf 3}_{L}$;

\item{10}: $K = Spin(6)$, $\vec
\CE \in \left( {\bf 6} \oplus \bar{\bf 6}
\right)_{r} \oplus {\bf 1}$

Alternatively,  in the last case we can use the octonionic representation with
$K = Spin(7)$, $\vec \CE \in {\bf
7}$;

We will simplify our integrals by deforming the
BRST operator. The deformation will involve
a choice of a generic element $\epsilon$ in
the  Cartan subalgebra of the
global symmetry group $K$.
We therefore choose elements
$\epsilon \in {\rm Lie} (Spin (2)),
 {\rm Lie}(Spin(4)),$ and ${\rm Lie}(Spin(6))$ for
$D=4,6,10$ respectively. Explicitly we will
write these elements as:
\eqn\css{\eqalign{& D = 4 \quad
\epsilon = \pmatrix{0 & E \cr -E & 0\cr}
   \cr
& D = 6  \quad  \epsilon = \pmatrix{\matrix{0 & E_{1} \cr -E_{1} &
0\cr} &
\cr & \matrix{0 & E_{2} \cr -E_{2} & 0\cr}\cr}   \cr
& D = 10  \quad \epsilon = \pmatrix{\matrix{0 & E_{1} + E_{2} \cr
-E_{1} - E_{2} & 0\cr} &   & \cr & \matrix{0 & E_{2} + E_{3} \cr
-E_{2}- E_{3} & 0\cr} & \cr
&  & \matrix{0 & E_{1} + E_{3} \cr -E_{1}- E_{3} & 0\cr}   \cr} \cr}}
for sufficiently generic real constants $E_i$.

Using the global symmetry one may deform the nilpotent charge \ssy\
to the differential of $K$-equivariant  cohomology:
\eqn\ssyi{\eqalign{Q_{\epsilon} X_{a} = \Psi_{a} \quad & Q_{\epsilon}
\Psi_{a} = [\phi , X_{a} ] + X_{b} T_{v}(\epsilon)^{b}_{a}\cr
Q_{\epsilon} \vec \chi = \vec H \quad & Q_{\epsilon} \vec H = [\phi, \vec \chi]
+
T_{s}(\epsilon) \cdot \vec \chi \cr
Q_{\epsilon} \bar\phi = \eta \quad & Q_{\epsilon} \eta = [\phi, \bar\phi] \cr
 Q_{\epsilon} & \phi = 0\cr }}
where we denote $T_{v}$ for the action of ${\rm Lie}(K)$ on $X$'s and
$T_{s}$ for the action of
${\rm Lie}(K)$ on the equations.
Now deform the action \rwraci\ to
\eqn\actnii{S_{\epsilon} = Q_{\epsilon} \left( {1\over{16 \tilde g}}
{\Tr} \eta[\phi, \bar\phi] - i {\Tr} \vec \chi \cdot \vec \CE +
g {\Tr} \vec \chi \cdot \vec H +
{1\over{4\hat g}} \sum_{a=1}^{D-2} {\Tr} \Psi_{a} [ X_{a}, \bar \phi]
\right)}
At this point the couplings $\hat g$, $\tilde g$ and $g$ are all
equal but in the sequel
we shall treat them separately. In particular we will
first take $\tilde g \rightarrow \infty$.
The new integral
$$
\int \ldots e^{-S_{\epsilon}}
$$
is convergent if the original \innt\ integral is convergent. In fact
the added piece
$S_{\epsilon} - S_{0}$ is equal to
$$
g {\Tr} \left( \vec \chi \cdot T_{s}(\epsilon) \vec \chi \right) +
{1\over{4\hat g}} T_{v}(\epsilon)^{ab}
{\Tr} \bar\phi [X_{a}, X_{b} ]
$$
which has ghost charge $-2$ (if we temporarily assign a charge zero
to $\epsilon$). This means that the
value of the integral (whose measure  has   net ghost charge zero) is
not
changed.
Now a closer look at the eigenvalues of $T_{s}$ reveals that there
is always one
zero eigen-value for the mass matrix of $\chi$
but the rest is non-vanishing for generic $\epsilon$.
We denote
this massless mode by    $\chi_{0}$, and
  consider adding to the action a $Q_{\epsilon}$-exact term
\eqn\mtr{
s Q_{\epsilon} {\Tr}(  \chi_{0} \bar\phi )}
with a large coefficient $s$. It has   ghost charge $-2$.
This term together with $g \vec H^{2}$ produces
masses for all the fermions of
negative ghost charge.
Integrating them out (by taking the limit $s
\to \infty$, $g \to \infty$) would produce
a very simple action but without a ``kinetic'' term for $\Psi_{a}$'s.
To cure this problem we add a
{\it positive} ghost charge operator
$$
{\half} t Q_{\epsilon} ( \sum_{i=1}^{D/2 - 1} B_{i}
\Psi_{i}^{\dagger} - B_{i}^{\dagger} \Psi_{i} )
$$
If we assign the standard ghost  charge $+2$ to $\epsilon$, then the
insertions of the coupling $t$ must be compensated by the
insertions of the coupling $s$, so the answer may only
depend on the combination
$st$. On the other hand, it is easy to repeat
the derivation of \witrevis\ by first taking the
limit $s \to \infty$ with  $g$ much smaller than $s$.
In this way one gets an effective action which is
schematically of the form:
\eqn\schemact{
S_{eff} \sim {1\over{s}}  \{ Q_{\epsilon} , {\Tr} {\Psi}_{a} [ X, \CE ] \}
}
and which has ghost charge two.
\foot{One might worry
that the original
integrals and the ones we are getting at this point differ
by   exponentially small terms, as   in \witrevis.
The difference with the situation
of \witrevis\ is that due to the absence of topologically non-trivial
solutions to the equations on finite-dimensional matrices there
are no extra contributions to the integral coming from infinity.}
As we shall see momentarily,  in the limit $s,t \to \infty$ the
dependence on
either variable actually vanishes, therefore
the value of integral which we get is equal to the original
integral \innt.

As discussed in \witrevis\krwjffr\higgs, one
can now proceed to do the integrals in the semiclassical
approximation for large $s,t,g$. We first do the Gaussian
integrals to eliminate the BRST quartet $(\eta, \bar\phi,
\vec \chi, \vec H)$. This results in a determinant in
the numerator of the measure of the form ${\rm Det}(\epsilon + ad(\phi))$
where the determinant is evaluated in the representation space
of the equations. Proceeding with the Gaussian integrals on
$(B_i, \Psi_i)$ produces determinants of the
form ${\rm Det}(\epsilon + ad(\phi))$ in the denominator. Finally,
taking into account the Vandermonde factor in reducing the
integral on $\phi$ from ${\rm Lie}(G)$ to ${\liet}=  {\rm Lie}(T)$ we obtain
the integral:
\eqn\fnli{\eqalign{
I_{D=10}(N) = &
\left(
{{(E_{1}+E_{2})
(E_{2}+E_{3})(E_{3}+E_{1})}\over{E_{1}E_{2}E_{3}E_{4}}}\right)^{N-1}
{{N}\over{N!}}
\int_{\liet} {\CD}\phi \prod_{i \neq j}
{{P(\phi_{ij})}\over{Q({\phi}_{ij})}} \cr
& P(x) = x (x + E_{1} + E_{2}) (x + E_{3} + E_{2}) (x + E_{1} + E_{3})
\cr
& Q(x) = \prod_{\alpha = 1}^{4} (x + E_{\alpha}+ i0)\cr}}
for the case $D=10$. Here $\sum_{\alpha} E_{\alpha} =0$
and the integral is taken along the real line.
Similarly, the same procedure gives the integral:
\eqn\fnlii{
I_{D=6}(N) =
\left( {{E_{1}+E_{2}}\over{E_{1}E_{2}}}\right)^{N-1} {{N}\over{N!}}
\int_{\liet} {\CD}\phi \prod_{i \neq j} {{\phi_{ij} (\phi_{ij} +
E_{1} + E_{2}) }\over{\prod_{\alpha = 1}^{2} (\phi_{ij} + E_{\alpha}+
i0)}}}
for $D=6$,
and can be obtained from \fnli\ by taking a {\it formal} limit $E_{3}
\to \infty$.
Finally, for $D=4$ the integral is:
\eqn\fnliii{
I_{D=4}(N) ={{N}\over{N! E_{1}^{N-1}}} \int_{\liet} {\CD}\phi \prod_{i
\neq j} {{\phi_{ij}  }\over{(\phi_{ij} + E_{1}+ i0)}}}
and can be obtained from \fnlii\ by taking a {\it formal} limit
$E_{2} \to \infty$.

The factor ${{N}\over{N!}}$ has the following origin. The denominator is
the order of the Weyl group of $SU(N)$ which
enters in passing to the integral over
the conjugacy classes of $\phi$. We then rewrite
this integral as an integral over $\liet$, divided by
$\vert W(G) \vert = N!$.  The numerator
$N$ is the order of the center of $\IZ_{N}$ which appears in
comparing the volumes of $SU(N)$ and $G$.
The measure
${\CD}\phi$ is defined as follows. The maximal Cartan subalgebra  of $SU(N)$
can be identified
with $\IR^{N-1}$ by means of the   imbedding:
$$
\left( \phi_{1} , \ldots , \phi_{N-1} \right) \to
{\rm diag} \left( \phi_{1} , \ldots , \phi_{N-1}, - \phi_{1}
- \ldots - \phi_{N-1} \right)
$$
into the space of traceless hermitian matrices. The measure
${\CD}\phi$ is simply the normalized Euclidean measure on $\IR^{N-1}$:
\eqn\msr{{\CD}\phi = \prod_{k=1}^{N-1} {{d\phi_{k}}\over{2\pi i}}
\quad . }

Finally, as mentioned in the introduction,
 it might appear that the integrals  \fnliii\fnlii\fnli\
are ill-defined since they are integrals along $\IR^{N-1}$
with a measure that generically approaches $1$ at
$\infty$. This is an illusion. They should be regarded
as contour integrals and become convergent once
a contour deformation prescription is adopted.
We will find such prescription in every case. The
 prescription
$E \to E + i0$ is required for the validity
of the Gaussian integrations, but we still
must give a prescription for closing the contours.
We  expect that the contour prescriptions found below
will follow  from a more careful implementation
of the technique of integrating out BRST quartets
than we have yet performed.

\newsec{Detailed evaluation for low values of $N$ }

\subsec{Two-body problem}

We begin by evaluating the integral \fnli\ for the
$D=10$ case:
\eqn\twonine{\eqalign{
I = &{1\over{2\pi i}} {{P^{\prime}(0)}\over{Q(0)}} \int_{\IR}  d\phi
{{P(2{\phi})P(-2{\phi})}\over{Q(2{\phi})Q(-2\phi)}}, \cr
& P(x) = x (x + E_{1} + E_{2}) (x + E_{3} +
E_{2})(x + E_{1} + E_{3})\cr
& Q(x) = {\prod}_{\alpha = 1}^{4} (x + E_{\alpha}+ i0)\cr}}

In order to evaluate it we close the contour in the upper half plane
(this is an example of the ``prescription'' alluded to above) and pick up
the contribution of four poles, at $\phi = {\half} E_{\alpha} +i0$.
The residue at $E_{\alpha}$ turns out to be
\eqn\rsd{{\rm Res}_{{\half} E_{\alpha}+i0} = {1\over{12}} {{R (-
2E_{\alpha})}\over{E_{\alpha} R^{\prime}(E_{\alpha})}}}
where
$$
R(x) = \prod_{\alpha=1}^{4} (x - E_{\alpha})
$$
and  the sum over the residues
can be evaluated using an auxiliary contour integral:
\eqn\evl{\sum_{\alpha=1}^{4} {\rm Res}_{{\half} E_{\alpha}} = {1\over{12}}
\left( \oint {{R(-2x)}\over{xR(x)}} dx - 1 \right) =
5/4}
For lower $D$'s the same formula \rsd\ holds, and the equation \evl\ gives
\eqn\lowerdee{
 {1\over{12}} \left( 2^{D/2-1}+(-1)^{D/2} \right)}
i.e. the famous $5/4$, $1/4$, $1/4$ for $D= 10, 6,4$
respectively originally computed in \pyi\sav.

\subsec{Three-body problem}

The formalism we have
developed so far is rather powerful. In fact,
it is still possible to evaluate the integral for $N=3$
directly. Let $x = \phi_{1} - \phi_{2}$, $y = \phi_{2} - \phi_{3} =
2\phi_{2} + \phi_{1}$.
The measure can be rewritten as:
$$
d\phi_{1} \wedge d\phi_{2} = {1\over{3}} dx \wedge dy
$$
Specializing \fnli\fnlii\ to this case we find
3 sets of possible poles. The first set is given by:
\eqn\setone{
x  \in \{ E_1+i 0, E_2+i 0, E_3+i 0, E_4+i 0\} \qquad
{\it and}
\qquad
y \in \{ E_1+i 0, E_2+i 0, E_3+i 0, E_4+i 0\},
}
The second set is
\eqn\settwo{
x \in \{ E_1+i 0, E_2+i 0, E_3+i 0, E_4+i 0\}
\qquad
{\it and}
\qquad
x + y \in \{ E_1+i 0, E_2+i 0, E_3+i 0, E_4+i 0\},
}
and the third set is:
\eqn\setthree{
x + y \in \{ E_1+i 0, E_2+i 0, E_3+i 0, E_4+i 0\}
\qquad
{\it and}
\qquad
y \in \{ E_1+i 0, E_2+i 0, E_3+i 0, E_4+i 0\} .
}
We order the $+i0$'s appropriately
so that ${\rm Im}(E_\alpha - E_\beta)>0$ for $\alpha > \beta$.
In the $D=6$ case
we have similar sets of poles but with $E_1,E_2$ present
 without   $E_3,E_4$.

In evaluating the integral we choose poles from the
first set but only take the second or third set (but not
both).  It is straightforward to evaluate the residues.
For example, for the 5+1 case
$x =E_1, y =E_1$ gives:
\eqn\conti{
({E_1+E_2 \over  E_1 E_2})^2
{E_1^2 E_2^2 (2E_1 + E_2)(3 E_1 +E_2)
\over  3 (E_1+E_2)^2 (E_1-E_2)(2 E_1-E_2)}
}
while the residue vanishes for $x = E_1, y =E_2$,
with a similar contribution with $1 \leftrightarrow 2$. Thus the sum of
the
first set of poles gives:
\eqn\contii{
({E_1+E_2 \over  E_1 E_2})^2
{2 E_1^2 E_2^2 (4 E_1^2 + 5 E_1 E_2 + 4 E_2^2)
\over  3 (E_1+E_2)^2 (E_1-2 E_2)(2 E_1-E_2)}
}
Choosing \settwo,  and not \setthree, the contribution
   $x + y =E_2,  x =E_1$  gives:
\eqn\contiii{
- ({E_1+E_2 \over  E_1 E_2})^2
{E_1^2 E_2^2 (2E_1 + E_2)( E_1 +2 E_2)
\over   (E_1+E_2)^2 (E_1-2 E_2)(2 E_1-E_2)}
}
The sum of \contii\ and \contiii\ is $1/3$, which leads
to $1/3^2$ for the net answer.
With a little more work one can check that in
the 9+1 case we obtain $Z= {1\over{3}} \left( 3 + 1/3 \right)$
(again, with $1/3$ coming from the factor ${{d\phi_{1} \wedge
d\phi_{2}}\over{dx
\wedge dy}}$).

\newsec{$SU(N)$ , $D=4$  }

For the $D=4$ case we may  use the
Bose-Cauchy identity:

\eqn\indty{ {1\over{E_{1}^{N}}} \prod_{i \neq j} {{\phi_{ij}
}\over{(\phi_{ij} + E_{1}+ i0)}} =
\sum_{\sigma \in \CS_{N}} (-1)^\sigma
\prod_{i=1}^{N} {1\over{\phi_{i} -
\phi_{\sigma (i)} + E_{1} +i0}} }

Of all the terms in \indty\ only the cycles of maximal
length $N$  can  contribute to
the residue evaluation (and there are $(N-1)!$ of those).
The integral \fnliii\ will pick up a residue for all $i$
except one (let us denote it by $j$)
provided that
$$
\phi_{\sigma(i)} = \phi_{i} + E_{1} + i0, \quad {\rm for \quad all
\quad } i \neq j.
$$
By relabelling the indices with the help of the Weyl group
we can assume that $j=N$ and
the permutaton $\sigma$ is a long
cycle $\sigma (i) = i+1$. The pole is at
\eqn\ple{\phi_{i} = {\half}( 2i - N -1) E_{1}}
and the residue is equal to: ${1\over{N^{2}}}$.
\foot{Notice that $\phi \in \liet$ can
be expressed as $\phi = \rho \cdot E_{1}$
where $\rho$ is half the sum of the positive roots.}

We prove this fact by taking the integral over the variables $\phi$
in the following order:
$\phi_{N-1}, \phi_{N-2}, \ldots, \phi_{1}$. In the sequel $E_{1}$
should read as $E_{1} + i 0$.

Given the fact that $\sigma = (1 2 3 \ldots N)$ we need to evaluate:
\eqn\intgr{\eqalign{& \qquad N {{E_{1} (-1)^{N-1}}\over{N (2\pi
i)^{N-1}}} \oint
\prod_{i=1}^{N-2} {{d\phi_{i}}\over{\phi_{i} -
\phi_{i+1} + E_{1}}}\cr
& {{d\phi_{N-1}}\over{
\left( 2\phi_{N-1} + \phi_{1} + \ldots + \phi_{N-2} + E_{1} \right)
\left( -2 \phi_{1} - \phi_{2} - \ldots - \phi_{N-1} + E_{1}
\right)}}\cr}
}
(the factor $N$ in the denominator
 is the order of the stabilizer of $\sigma$ in
the Weyl group: $N!/ (N-1)!$ and the sign $(-1)^{N-1}$ is
$(-1)^{\sigma}$
for the long cycle).
Let us  prove by induction that the integral \intgr\ reduces to
\eqn\iii{
\eqalign{
{{k E_{1} (-1)^{N-k}}\over{(k+1)^{2}  (2\pi i)^{N-k}}}
& \oint  \prod_{i=1}^{N-k-1} {{d\phi_{i}}\over{\phi_{i} -
\phi_{i+1} + E_{1}}} \cr}
}
$$
{{d\phi_{N-k}}\over{
\left( \phi_{N-k} + {1\over{k+1}} \left( \phi_{1} + \ldots +
\phi_{N-k-1}
\right)  + {{k}\over{2}} E_{1}  \right)
\left( - \phi_{1} - {1\over{k+1}}\left( \phi_{2} + \ldots + \phi_{N-k}
\right)  +
{{k}\over{2}}
E_{1}\right)}}
$$
For $k=1$ this expression is identical to \intgr. Now let us take the
$\phi_{N-k}$ integral. By closing the contour in either the upper or
the lower half plane (it doesn't matter) we pick up either
one or two residues. For
simplicity we always close the integral in the
lower half-plane, meaning that:
\eqn\iv{
\phi_{N-k} = - {{k}\over{2}} E_{1} - {1\over{k+1}} \left( \phi_{1}
+ \ldots + \phi_{N-k-1} \right)
}
By evaluating the residue we immediately see that the declared
form of the integral is reproduced with the replacement
$k \to k +1$.
Finally, for $k = N-1$ we get
\eqn\ivi{
{{E_{1}(N-1) (-1)}\over{N^{2} 2\pi i}} \oint
{{d\phi_{1}}\over{
\left( \phi_{1} +  {{N-1}\over{2}} E_{1}  \right)
\left( - \phi_{1} + {{N-1}\over{2}}
E_{1}\right)}} = {1\over{N^{2}}}
}
Hence, the $D=4$ integral is equal to
\eqn\ivii{
I_{D=4}(N) = {1\over{N^{2}}}
}

Note that the integral has been localized to the fixed point of the
$\IC^{*}$
action on the quotient of the space of regular  traceless matrices $B$ by the
adjoint
action
of the group $SL_{N}({\IC})$.
\foot{
A group element is {\it regular} if its   centralizer
in $SL_{N}({\IC})$  has
dimension $N-1$.}
Indeed, the $\phi$ from \ple\ solves
the
equation
\eqn\fxpnt{[ B, \phi] = E_{1} B}
for $B_{ij} = \delta_{i, j-1}$. On general principles we
expect the integral to localize to the
$Q_\epsilon$ fixed-points.
Of course, the equation \fxpnt\ has other,
more non-trivial, solutions.
In fact, for every Jordan cell decomposition
\eqn\jordcell{
B = \pmatrix{J_{1} & 0 & 0 \cr
0 & \ldots & 0 \cr
0 & 0 & J_{k} \cr}
}
for $J_{l}$ being a Jordan block of length $n_{l}$, $\sum_{l} n_{l} =
N$
we get a solution to \fxpnt\  of the form:
\eqn\jordcellp{
\phi = \pmatrix{\varphi_{1} & 0 & 0 \cr
0 & \ldots & 0 \cr
0 & 0 & \varphi_{k} \cr}
}
where $\varphi_{l} = f_{l} {\Id}_{n_{l}} + {\rm diag} \left( {\half}(
2i - n_{l} -1) E_{1} \right)$,
$i= 1, \ldots, n_{l}$. The parameters $f_{l}$ are only constrained by
the
requirement that ${\Tr}\phi =0$, which leaves $k-1$ free zero modes.
But the
presence of extra zero modes is equivalent to the statement that the
integrand in
\fnliii\ can't  pick up sufficiently many  residues. Before
eliminating the redundant fields every mode of
$\phi$ came together with a bunch of superpartners, fermionic
modes among them. By supersymmetry the unlifted modes,
the
$f_{l}$'s,  correspond to the
extra fermionic modes which make the integral vanish.
We thus obtain the following important principle:
{\sl The fixed points with extra $U(1)$'s left  unbroken don't
contribute to the index.}
It is interesting to compare this principle with the one derived in
\estring\ in a seemingly different context.

\newsec{$SU(N)$ ,  $D=6$ }

In this case we rewrite the integral \fnlii\ as:
\eqn\fv{
{1\over{(N-1)!}} \left( {{E_{1}+E_{2}}\over{E_{1}E_{2}}}\right)^{N-1}
{1\over{(2\pi i)^{N}}} \oint {{d\phi_{1}
\wedge \ldots \wedge d\phi_{N}}\over{\phi_{1} + \ldots + \phi_{N}}}
\prod_{i \neq j} {{\phi_{ij} (\phi_{ij} + E_{1} + E_{2})
}\over{\prod_{\alpha = 1}^{2} (\phi_{ij} + E_{\alpha}+ i0)}}}
We next perform the change of variables:
\eqn\cnv{\phi_{i} \mapsto \tilde\phi_{i} = \phi_{i} +
\sum_{j=1}^{N-1}
\phi_{j}  \qquad i = 1, \dots, N . }
The measure gets an extra factor ${1\over{N}}$:
\eqn\msr{{{d\phi_{1} \wedge \ldots \wedge d\phi_{N}}\over{\phi_{1} +
\ldots + \phi_{N}}} =
{1\over{N}}{{d\tilde\phi_{1} \wedge \ldots
\wedge d\tilde\phi_{N}}\over{\tilde\phi_{N}}} }
and we may rewrite \fv\ as:
\eqn\fvi{\eqalign{&
{{(E_{1}+E_{2})^{N-1}}\over{N (2\pi i)^{N} (E_{1}E_{2})^{N-1}}}  \oint
{{d\tilde\phi_{1}
\wedge \ldots \wedge d\tilde\phi_{N}}\over{\tilde\phi_{N}}} \times
 \prod_{i \neq j} {{\phi_{ij} (\phi_{ij} + E_{1} + E_{2})
}\over{\prod_{\alpha = 1}^{2} (\phi_{ij} + E_{\alpha}+ i0)}} = \cr&
{{E_{1}E_{2}}\over{N (2\pi i)^{N} (E_{1}+ E_{2})}}  \oint
d\phi_{1}
\wedge \ldots \wedge d\phi_{N}
 \prod_{i < N} (- \phi_{i})  \prod_{i} (\phi_{i} + E_{1} +E_{2})
\times \cr & \times
{{\prod_{i \neq j} {\phi}_{ij} }\over{\prod_{i}
(-\phi_{i}) (\phi_{i} + E_{1} +E_{2})}}
\prod_{i,j} {{(\phi_{ij} + E_{1} + E_{2})}\over{(\phi_{ij} +
E_{1})(\phi_{ij} + E_{2})}}\cr}}
where in the second line we made a substitution: $\tilde\phi \to \phi$
and in the denominators $E_{\alpha} \to E_{\alpha} + i0$.
The factor $(N-1)!$ disappears for the following reason. The choice of
$\phi_{N}$ breaks the permutation group to $\CS_{N-1}$. We can fix
the latter symmetry by ordering the eigenvalues $\phi_{i}$. As we shall
see later, in assigning the poles of the integral \fv\ to   Young
tableaux
each tableau yields a definite way of ordering the eigenvalues
which takes up the whole of $\CS_{N-1}$.

Despite the seemingly senseless
manipulation we have arrived at an integral we can make sense of and in fact
evaluate.
In order to explain its meaning we recall that the solutions to
the equation $[ B_{1}, B_{2} ] =0$ modulo
conjugation describe the symmetric product of $\IC^{2}$ away from singularities
and in fact provide a certain resolution of singularities, once
appropriate stability conditions are imposed. These stability
conditions can be formulated by introducing an auxiliary vector
$I \in \IC^{N}$. Then the stable data consists of a triple
$Z = (B_{1}, B_{2}, I)$, such that $[B_{1}, B_{2}]=0$ and
there is no proper $B_{1}, B_{2}$ invariant subspace of $\IC^{N}$
which   contains $I$. The triples
$(B_{1}, B_{2}, I)$ and $(g^{-1} B_{1} g, g^{-1}B_{2} g, g^{-1} I)$
are considered equivalent for any $g \in {\rm GL}_{N}({\IC})$.
It can be shown that the equivalence classes of
such data $Z$  are in one-one correspondence
with  codimension $N$ ideals
 ${\CI}_{Z}$  in ${\IC} [ z_{1}, z_{2}]$.
\foot{Briefly, $V_Z\equiv {\IC} [ z_{1}, z_{2}]/{\CI}_{Z}$ is an
$N$-dimensional complex vector space.
The linear operators $B_1, B_2$ are the operations
of multiplication by $z_1, z_2$, respectively,
projected to endomorphisms of $V_Z$. The vector
$I$ is the image of $1\in \IC[z_1, z_2]$. The
inverse map proceeds by identifying the span
of $\{ B_1^n B_2^m \cdot I\}_{n,m\geq 0}$
with $V_Z$. This is explained in details 
in Theorem 1.14, page 10, of \nakheis.}
The set of all codimension $N$ ideals in the
polynomial ring $\IC[z_1,z_2]$
 forms what is called the ``Hilbert scheme
of $N$ points on $\IC^{2}$,''  and is
denoted by  $\CH_{N} = \left(
{\IC}^{2}\right)^{[N]}$.
The quotients $V_Z = \IC [ z_{1}, z_{2} ] /{\CI}_{Z}$
are the fibers of  a rank $N$ vector bundle $\CE$ over $\CH_{N}$.
The Chern roots of $\CE$ are nothing but $- \phi_{i}$'s.
The space $\CH_{N}$ is acted on by the complex torus
${\bf T} = \IC^{*} \times  \IC^{*}$
by rotation of  the coordinates $(z_{1}, z_{2})$:
\eqn\rota{
(z_{1}, z_{2}) \mapsto (e^{E_{1}} z_{1}, e^{E_{2}} z_{2}).
}
This action lifts to the action on the data $(B_{1}, B_{2}, I)$
as follows:
\eqn\rotap{
(B_{1}, B_{2}, I) \mapsto ( e^{E_{1}} B_{1} , e^{E_{2}} B_{2}, I)
}
The action of $\bf T$ on $\CE$ is defined through the identification
of the fiber $\CE_{(B_{1}, B_{2}, I)}$ with
the vector space $\IC [ B_{1}, B_{2} ] I$.
Let $Q$ be the
topologically trivial  $\bf T$-equivariant rank $2$ vector bundle
over
$\CH_{N}$ whose  isotypical decomposition
  coincides with that of the space $\IC^{2}$
with coordinates $z_{1}, z_{2}$.
The integral \fvi\ computes the Euler character
of a certain $\bf T$-equivariant  bundle $\CF_{N}$ over $\CH_{N}$.To be more
precise,   we need the virtual bundle
given by:
\eqn\virtbundle{
\CF_N=
\left( Q \oplus \CE \oplus \CE^{*}\otimes \wedge^{2}Q \right) \ominus
\left( {\rm det}\CE
 \oplus  \wedge^{2}Q\right).}
This bundle has virtual dimension $2N$.
 The Euler classes of the
various factors can be recognized in the integrand of
\fvi. For example, the Euler class of
$\CE^{*}\otimes \wedge^{2}Q$ is the  product
$\prod_i (\phi_i  +E_1 + E_2)$, while the 
incomplete product
$\prod_{i<N} (-\phi_i)$ gives, roughly speaking, the  class of
$\CE - {\rm det}\CE$.
The factors involving
$Q$ lead to the overall factors involving $E_i$,
and the third line of \fvi\ is a measure factor
for integration over $\CH_N$.

The evaluation of \fvi\ by residues is
equivalent to the use of fixed point
techniques (see \kirwan\krwjffr\
for more examples of such techniques).
We   now make a slight detour and remind the
reader of
the ideology behind  such computations \atbott\witrevis.
Suppose one wishes to compute the integral
$$
\int e^{-{{S}\over{\hbar}}} DX
$$
in the quasiclassical approximation $\hbar \to 0$. In general
one has to take into account
a certain set of critical points of $S$ and include the determinants
of the matrix of second derivatives of $S$. Some integrals
have the property of having exact quasiclassics. One should take
into account all critical points of $S$ and compute the determinants
which would have in general $\pm 1$ signs for unstable critical points.
One famous example of such an  integral is the Duistermaat-Heckmann formula:
\eqn\dhform{
\int_{M^{2m}} {{\omega^{m}}\over{m!}} e^{-tH} =
\sum_{p: dH (p) = 0} {{e^{-tH(p)}}\over{\prod_{i=1}^{m} t  m_{i}(p)}}
}
where $(M, \omega)$ is a symplectic manifold with Hamiltonian  $U(1)$
action generated by $H$, $p$'s are the fixed points of the $U(1)$
action
(assuming they are isolated) and $m_{i}(p)$ are the weights of
the $U(1)$ action in the tangent space to $M$ at the fixed point $p$.
Of course, there exist generalizations of this formula
for other manifolds, groups other than  $U(1)$, non-isolated fixed points
and so on.
In the problem of present interest it turns out that
the fixed points
are enumerated by   Young tableaux $D$  with $\# D = N$ boxes. \foot{We
thank
V.~Ginzburg for very clear explanation of this fact. In the language
of ideals $\CI_{Z}$ the fixed points are the
ideals which are spanned by $z_{1}^{a} z_{2}^{b}$
with $a \geq \nu_{b}, b \geq \nu_{a}^{\prime}$\ginz. It explains the
formula
for the weights $\phi_{(\alpha, \beta)}$ below.}
In other words, consider the partition $N = \nu_{1} + \ldots +
\nu_{\nu_{1}^{\prime}}
= \nu_{1}^{\prime} + \ldots + \nu_{\nu_{1}}^{\prime}$.
Let $(\alpha, \beta)$ denote the position of a box in the Young
tableau. There is  the  one-to-one correspondence between the
labels $i \in \{ 1, \ldots, N \}$
and the allowed pairs $(\alpha, \beta)$:
$1 \leq \alpha \leq \nu_{\beta}$, $1 \leq \beta \leq
\nu^{\prime}_{\alpha}$, given by the lexicographic order
($(\alpha, \beta) > (\alpha^{\prime}, \beta^{\prime})$ if
$\alpha < \alpha^{\prime}$ or $\beta < \beta^{\prime}$ for
$\alpha = \alpha^{\prime}$). In particular,  $(1,1) \leftrightarrow N$.
The corresponding eigenvalues $\phi_{i}$ are given by:
\eqn\ffi{\phi_{(\alpha, \beta)} = ({\alpha}-1) E_{1} + ({\beta} - 1)
E_{2}}
One can evaluate the residue at \ffi\ using the results of \nakheis.
Namely, in  \nakheis\
 it is proven that for a Young tableau
$D$ and the set $\phi_{i}$ given by \ffi\ the following sum:
\eqn\smr{\sum_{i,j \in D} \bigl[
e^{\phi_{ij}} + e^{\phi_{ij} + E_{1} + E_{2}}
-
e^{\phi_{ij} + E_{1}} - e^{\phi_{ij} + E_{2}} \bigr] -
\sum_{i \in D} \bigl[ e^{-\phi_{i}} + e^{\phi_{i} + E_{1} + E_{2}}\bigr] }
is equal to:
\eqn\smmr{- \sum_{(\alpha, \beta) \in D} e^{( \nu_{\beta} - \alpha + 1)
E_{1} + (\beta -
\nu_{\alpha}^{\prime})E_{2}} + e^{(\alpha - \nu_{\beta}) E_{1} +
(\nu_{\alpha}^{\prime} - \beta
+1) E_{2}}}
In fact, in \nakheis\ the weight decomposition of the
tangent space to $\left( \IC^{2} \right)^{[N]}$
at the fixed point corresponding to $D$ is computed. It is encoded in
the
formula \smmr. We simply have to take
the product of those weights, which will go into the
denominator. In addition we need to take into
account the decomposition of the bundle $\CF_{N}$
into weight subspaces and compute the
product of those weights, which will go into the numerator.
We simply use the fact that
the weights of $\CE$ are given by $\phi_{i}$'s.
Combining these two products we arrive at:
\eqn\yng{\eqalign{ & Y_{D} \equiv {\rm contribution \quad of \quad }
D = \cr
(-)^{N-1} E_{1}E_{2} &
{{\prod_{(\alpha, \beta) \neq (1,1)}
\left(  ({\alpha}-1) E_{1} + ({\beta} - 1) E_{2} \right)
\left( \alpha E_{1} + \beta E_{2} \right)}\over{
\prod_{(\alpha, \beta)}
\left( ( \nu_{\beta} - \alpha + 1) E_{1} + (\beta -
\nu_{\alpha}^{\prime})E_{2}\right)
\left((\alpha - \nu_{\beta}) E_{1} + (\nu_{\alpha}^{\prime} - \beta
+1) E_{2}
\right)}}\cr}
}
Now what remains is to sum over all Young tableaux $D$.

One can check that the explicit pole prescriptions
found above for the $SU(2), SU(3)$ cases are reproduced
by the poles associated to Young diagrams. Moreover,
as a further illustration (and to have a look  at the case with $N$ non-prime)
we write out all the residues for the $SU(4)$ case: there are five Young
tableaux,
a column $(4)$, a hook $(3,1)$, a box $(2,2)$, the mirror hook $(2,1,1)$
and a row $(1,1,1,1)$ (in the brackets
we listed the values of $\nu_{\alpha}$'s).  Let $x = E_{2}/E_{1}$.
The contributions are:
\eqn\clmn{\eqalign{& \matrix{{\rm column} & (4) & - {1\over{4}}
{{(1+2x)(1+3x)(1+4x)}\over{(1-x)(1-2x)(1-3x)}} \cr
{\rm hook}&  (3,1) &
- {1\over{2}} {{(1+2x)(1+3x)(x+2)}\over{(1-x)^{2}(-1+3x)}} \cr
{\rm box}& (2,2) &  - {1\over{2}}
{{(1+2x)(x+2)(1+x)^{2}}\over{(1-2x)(2-x)(1-x)^{2}}} \cr
{\rm hook} & (2,1,1) & - {1\over{2}} {{(1+2x)(x+3)(x+2)}\over{(1-x)^{2}(-x+3)}}
\cr
{\rm row}&  (1,1,1,1) & - {1\over{4}}
{{(x+2)(x+3)(x+4)}\over{(x-1)(x-2)(x-3)}} \cr}\cr
& {\rm the \quad sum} \qquad\qquad\qquad\qquad {1\over{4}}\cr}}
which together with the $1/4$ factor from the measure gives $1/16$ as the
answer.

The general answer is also expected to be $E_1, E_2$ independent.
Looking at \yng\ we see that the factors which contain single $E_1$'s
cancel out. Indeed, in the numerator these come from
$\beta =1$ in the factors
$ ({\alpha}-1) E_{1} + ({\beta} - 1) E_{2}$,
producing
\eqn\numr{E_1^{\nu_1} (\nu_1 - 1)!}
In the denominator
the single $E_1$'s come from $\beta = \nu_{\alpha}^{\prime}$
in the factors $( \nu_{\beta} - \alpha + 1) E_{1} + (\beta -
\nu_{\alpha}^{\prime})E_{2}$, giving rise to the product:
\eqn\dnm{E_{1}^{\nu_{1}} \prod_{\alpha =1}^{\nu_{1}} \left
( \nu_{\nu_{\alpha}^{\prime}} - {\alpha} +1 \right) =
E_{1}^{\nu_{1}} ({\nu}_{1}-1)!}
Hence, single $E_{1}$'s cancel out and the limit $E_{1} \to 0$ is
well-defined.
It is easy to see  that all other factors cancel out except for the
overall sign $(-)^{\nu_{1}-N}$, coming from comparing
the products $\prod_{\beta < \nu_{\alpha}^{\prime}} ( {\beta} -
{\nu}_{\alpha}^{\prime} ) $ and $\prod_{\nu_{\alpha}^{\prime} \geq
\beta >
1}
(\beta - 1)$.
Thus we are left with:
\eqn\smlfcntr{Y_{D} = {{(-)^{\nu_{1}-1}}\over{N}}
{{(\nu_1-1)!}\over{\prod_{\alpha} \left
( {\nu}_{\nu_{\alpha}^{\prime}} - {\alpha} +1 \right)}}}
Scary as it seems, the expression \smlfcntr\ can be represented in a
very simple form. The way to do it is to combine the factors in the
denominator into the groups with constant $\nu_{\alpha}^{\prime}$.
A little mental excercise shows that the result can be represented as
follows:
\eqn\rslt{Y(q) =
\sum_{D}q^{\# D}  Y_{D} = \sum_{\{ \ell_\gamma \}, \sum_\gamma
\ell_\gamma >0, \gamma = 1,2,\ldots, \ell_{\gamma} \geq 0 }
{{q^{\sum_{\gamma} {\gamma}\ell_{\gamma}}}\over{\sum_{\gamma}
\gamma {\ell_{\gamma}}}}
(-)^{\sum_{\gamma} \ell_{\gamma}} {{\left(
\sum_{\gamma}
\ell_{\gamma} - 1 \right)!}\over{\prod_{\gamma} \ell_{\gamma} !}} }
Here $\ell_{\gamma}$ represent yet another way of partition $N$ into
the
sum of positive integers:
$$
N = \sum_{\gamma=1}^{\infty} \gamma \ell_{\gamma}
$$
and $\ell_{\gamma} = \# \{ \alpha \vert \nu_{\alpha}^{\prime} =
\gamma\}$,
in particular $\nu_{1} = \sum_{\gamma} \ell_{\gamma}$.
The rest is easy: represent the factorial in the numerator of \rslt\
and $\sum_{\gamma} \gamma \ell_{\gamma}$ in the denominator
with the help of integrals:
\eqn\rslti{\eqalign{
Y(q) = -\int_{0}^{\infty} ds \int_{0}^{\infty} {{dt}\over{t}} e^{-t}
\sum_{\{ \ell_{\gamma} \}} &
\prod_{\gamma=1}^{\infty}
{{\left( - t q^{\gamma} e^{-s \gamma}
\right)^{\ell_{\gamma}}}\over{\ell_{\gamma}!}} = \cr
= -\int \int_{0}^{\infty} {{ds dt}\over{t}} e^{-t} \left(
e^{-t{{qe^{-s}}\over{1-qe^{-s}}}} -
1\right) = &
- \int\int_{0}^{\infty} {{ds dt}\over{t}}
\left( e^{-{{t}\over{1-qe^{-s}}}} - e^{-t}\right) = \cr
- \int ds {\rm log} (1-qe^{-s}) =
{\rm Li}_{2}(q) = \sum_{N=1}^{\infty} {{q^{N}}\over{N^{2}}} \cr}}
So we get:
\eqn\deesixans{
I_{D=6}(N) = {1\over{N^{2}}}
}
just as in the $3+1$ case.

It is probably worth pointing out that the last stage
of computations is very similar to those performed in \manin\
in the course of proving that the contribution to a prepotential
of an isolated
rational curve sitting in Calabi-Yau manifold  equals ${\rm Li}_{3}(q)$.

Another important remark is that a faster way of getting the
equality $I_{D=6} = I_{D=4}$ is by taking the limit $E_{2} \to \infty$.
One might also attempt to take the limit $E_{3} \to \infty$ in the
$D=10$ integral. This needn't (and in fact doesn't) work
because the sum rule $\sum E_\alpha=0$ then
forces  $E_{4} \sim - E_{3} \to
\infty$ too, and the contour integration is ``pinched'' between the poles.
Pinching poles in a contour integral is a well-known
source of discontinuity.

\newsec{$SU(N)$, $D=10$  }

This section concludes our tour of the matrix integrals.
In principle the integral \fnli\ may be computed by
summing over a set of generalized Young tableaux
(as we did above for $N=2,3$).
It turns out, however,
that there is a shorter route to the answer, which avoids
working with  any new integrals. The strategy is to reduce the
number of matrices by enforcing  deformed octonionic
instanton equations.
As opposed to   section $2$ where we were basically taking strong
coupling limits  here we are taking mixed weak and strong
coupling limits, imposing the weak coupling limit to enforce
some of the equations.

Let us take $E_{\alpha} =0$ for all $\alpha$.
Introduce the formal variable $m$.
Consider the expression
\eqn\neqn{\Phi_{ij} = [ B_{i}, B_{j} ] - m \epsilon_{ijk4} B_{k} \qquad\qquad
1\leq i,j\leq 4 }
The
instanton equations  may now be deformed to
\eqn\nwisnt{\CE_{ij} = \Phi_{ij} -{\half}  \epsilon_{ijkl}
\Phi_{kl}^{\dagger}}
Note that
\eqn\idn{
{\half} \sum_{1\leq i,j\leq 4} {\Tr} \CE_{ij} \CE_{ij}^{\dagger} =  \sum_{1\leq
i,j\leq 4} {\Tr}
\Phi_{ij} \Phi_{ij}^{\dagger}  }
Hence  the equations $\CE_{ij} = 0$ imply:
\eqn\vcua{[B_{i} , B_{j} ] = m \epsilon_{ijk4} B_{k}, }
\eqn\vcuua{ [B_{4} , B_{k} ]
= 0.}
The equations \vcua\ are formally
the equations for the vacua of $\CN=4$ broken
down to $\CN=1$ (see \vw). Equation \vcuua\ implies
that $B_{4}$ generates the gauge transformations in the
complexified unbroken group.

Now let us  take  separate couplings
$g^{\prime}, g^{\prime\prime}$
for the equations
$\CE_{ij}$ and for the equation
$\sum_{i=1}^{4} [B_{i}, B_{i}^{\dagger}]$ respectively
(we can do this without spoiling $Q$-symmetry). Take the limit
$g^{\prime} \to 0$. This limit  enforces   equations \vcua\vcuua.
We also split the coupling ${1\over{\hat g}} {\Tr} [X_{a}, \phi]^{2}$
as follows:
\eqn\split{
{1\over{\hat g}} \sum_{a=1}^{4}{\Tr} \vert [X_{a}, \phi] \vert ^{2} \to
{1\over{\hat g^{\prime}}} \sum_{i=1}^{3} {\Tr} \vert [B_{i}, \phi] \vert^{2}
+ {1\over{\hat g^{\prime\prime}}}  {\Tr} \vert [B_{4}, \phi] \vert^{2}}
Upon taking the limit ${\hat g}^{\prime} \to 0$ we enforce the equations  
$[B_{i} , \phi] = 0, i=1,2,3$.

Adopting the argument
that extra $U(1)$'s kill the contributions to the
partition function we only have to count
the vacua where the adjoint gauge group is broken down
to $SU(d)/{\IZ_{d}}$, for $N = ad$.
For these vacua:
\eqn\vcua{B_{\alpha} = \Vert L_{\alpha} \Vert_{a \times a}
\otimes {\Id}_{d \times d}}
for $\alpha = 1,2,3$, $L_{\alpha}$ being $SU(2)$ generators in the
$a$-dimensional irreducible representation of $SU(2)$.
Also, we have
\eqn\unbrk{\left( B_{4} \right)_{N \times N} = {\Id}_{a \times a}
\otimes \left( B_{4} \right)_{d \times d}, \quad
\left( \phi \right)_{N \times N} = {\Id}_{a \times a}
\otimes \left( \phi \right)_{d \times d}.}
In the limit we are taking we can integrate out
the $B_\alpha, \alpha=1,2,3$ degrees of freedom,
leaving behind $B_4, \phi$.
Accordingly, we recognize that we have
exactly the degrees of freedom present in
the integral $I_{D=4}(d)$ for gauge group
$SU(d)/{\IZ_{d}}$. Moreover,  due to supersymmetry,
not only the degrees of freedom but also the
measure is appropriate to interpret
 the integral as $I_{D=4}(d)$. Now, we showed above
that $I_{D=4}(d)= 1/{d^2}$.
Thus, we conclude that the answer
is:
\eqn\aswr{ I_{D=10}(N) = \sum_{d \vert N} {1\over{d^{2}}} }
and in particular is equal to  $1+1/N^2$  only for $N$ prime.
The term with $d=1$ comes from the vacuum with
completely broken gauge group.

\newsec{Comparison with partition
functions of susy gauge theory
on $T^{4}$ and $K3$}

There are some interesting relations of the
integrals $I_D(N)$ with other well-studied partition
functions. First there is a relation with $5$-branes.
It is worth noting that the $q^0$ term in the
partition function of   $N$ fivebranes wrapped on $K3$
proposed in \estring\ reproduces the
answer \aswr\ for all $N$. One must divide by
$24$, which is the Euler characteristics of the moduli space
of a center of mass of  $D0$ branes moving on $K3$.
The partition function is
computed by wrapping the worldvolume of the fivebranes on $K3 \times
T^{2}$,
which by a series of $T$- and $S$-dualities can be mapped
to the problem of $N$ $D4$-branes wrapped on $K3$
and $N$ $D0$ branes bound to it. The $q^{0}$ term counts the zero
$D0$-brane charge sector in the effective gauge theory. Presumably,
by a Fourier-Mukai-Nahm-duality of $K3$ surface one can map this problem
to the problem of $N$ $D0$ branes in ten dimensions, by taking the limit
of very large $K3$ surface on which $N$ $D0$ branes propagate.

A more direct connection is that between
$I_D(N)$ and partition functions of $SYM$ on tori.
Consider $SU(N)/{\IZ}_{N}$ $\CN=4$ SYM on $T^{4}$, viewed as the theory
of
$N$ $D3$-instantons wrapped on $T^4$ with the center of
mass motion factored out (otherwise the partition function
vanishes). Again, the mass perturbation breaks the theory to $\CN=1$
with unbroken groups without $U(1)$'s being\foot{We thank C.~Vafa
 for the clarifying discussion on this point}
\eqn\unbrk{SU(d)/{\IZ}_{d}, \quad ad = N}
The $SU(d)$ $\CN=1$ theory has $d$ vacua, each contributing
$1$ to the partition function and their total contribution
is $d$. The partition function of $\CN =1$ $SU(d)/{\IZ}_{d}$
theory is $d^{3}$ times smaller, since the partition function
of $SU(d)$ contained as a factor the number of $\IZ_{d}$ flat
connections ($d^{4}$) and the volume of $SU(d)/{\IZ}_{d}$ is $d$
times smaller, see \vw\ for more detailed explanations.
Hence the partition function of $\CN =1$ $SU(d)/{\IZ}_{d}$
gauge theory on the four-torus
\foot{In the zero 't Hooft
magnetic flux sector}\  is equal to
\eqn\fourtor{
{1\over{d^{2}}}
}
and the partition function of the $SU(N)/{\IZ}_{N}$ $\CN =4$ theory is
given by:
\eqn\prtn{{\CZ}_{SU(N)/{\IZ}_{N}}^{{\CN}=4}(T^{4}) =
\sum_{d \vert N} {1\over{d^{2}}} }
Then the $T$-duality presumably 
relates the partition function of $N$ $D3$ branes
wrapped on $T^4$ to that of $D(-1)$ instantons in ten dimensions.
This concludes the proof of the conjecture of  \grn.

For   lower numbers of supersymmetries the partition functions
of the $SU(N)/{\IZ}_{N}$ are easy. For $\CN=1$ as we argued we get
${1\over{N^{2}}} = N / N^{3}$, where $N$ in the numerator
is Witten's index \witconst\ and the factor $N^{3}$ is the
effect of the center $\IZ_{N} \subset SU(N)$. For $\CN=2$, standard lore says
that by the mass perturbation the theory
reduces to $\CN=1$ and this perturbation does not affect the value
of the partition function \Witfeb.
So, we get:
\eqn\parti{
\CZ_{SU(N)/{\IZ}_{N}}^{\CN = 1,2} (T^{4}) = {1\over{N^{2}}}
}
For the minimal supersymmetric three-dimensional gauge theory
with the gauge group $SU(N)$ Witten's index is equal to $1$. The effect
of flat $\IZ_{N}$ connections is now $N^{3-1} = N^{2}$ thus leading
to the same answer
\eqn\partii{
\CZ_{SU(N)/{\IZ}_{N}}^{\CN= 1,2} (T^{3}) = {1\over{N^{2}}}
}
One could also get this answer by adding a Chern-Simons term
to the SYM Lagrangian (suitably accompanied by the fermions
so as to preserve some susy, see \nikitafive\cs) and then analytically
continuing in $k$ - the coefficient in front of the CS term.
It would be
interesting to see whether the above  answer could be reproduced by the
finite dimensional integral of the sort we have considered in the paper.
As has been pointed out in \sav\ for even $N$ and subsequently
argued  in \nicolai\ for all $N$
the $D=3$ integral should vanish. The reason (at least for even $N$)
being that the fermionic Pfaffian is odd under the parity
reversal $X \to - X$. On the other hand, by adding the Chern-Simons-like
term:

$$
k \left( {\Tr} X [ \phi, \bar\phi] + \psi\eta \right)
$$
and integrating out all massive modes we arrive
at the integral of the same form as the one for $D=4$, which
should be equal to ${1\over{N^{2}}}$
thus providing an agreement with the field theory
computation. Clearly, this CS-like
term violates parity. On the other hand, the original integral
is not obviously absolutely convergent, therefore the parity
arguments may be invalid. It would be interesing to resolve this
puzzle.

Another interesting question, but one which is   beyond the scope of this
paper,
is
the applications to the IKKT model \ikkt. In fact, our technique
allows for the derivation of regularized correlation    functions of
the operators ${\Tr} \phi^{n_{1}} \ldots {\Tr} \phi^{n_{k}}$.

\newsec{Acknowledgements}

We would like to thank
T.~Banks, L.~Baulieu, M.~Green,  S.~Sethi, I.~Singer, M.~Staudacher and
C.~Vafa
for  useful remarks and discussions.
G.~M. and N.~N. are grateful to the ITP at Santa Barbara and
especially to D.~Gross for hospitality
and to the organizers and
participants
of the Workshops on Geometry and String Duality for
providing a stimulating atmosphere.
S.~Sh. is grateful to the Theory Group at CERN for hospitality.

 The research of G.~Moore
is supported by DOE grant DE-FG02-92ER40704,
 that of S.~Shatashvili,
by DOE grant DE-FG02-92ER40704, by NSF CAREER award and by
OJI award from DOE and by Alfred P.~Sloan foundation.
The research of N.~Nekrasov was supported by Harvard Society of
Fellows,
partially by NSF under  grant
PHY-92-18167, partially by RFFI under grant 96-02-18046 and partially
by grant 96-15-96455 for scientific schools.

In addition, this research  was supported in part by NSF
under Grant No. PHY-94-07194.

\listrefs
\bye